\documentclass[preprint]{aastex61}
\usepackage{amssymb,amsmath}
\usepackage{graphicx}
\usepackage{natbib}
\usepackage[OT1,T1]{fontenc}

\begin{document}

\title{The Response of the Lyman-Alpha Line in Different Flare Heating Models}

\author{Jie Hong}
\affiliation{School of Astronomy and Space Science, Nanjing University, Nanjing 210023, China}
\affiliation{Key Laboratory for Modern Astronomy and Astrophysics (Nanjing University), Ministry of Education, Nanjing 210023, China}

\author{Ying Li}
\affiliation{Key Laboratory of Dark Matter and Space Astronomy, Purple Mountain Observatory, Chinese Academy of Sciences, Nanjing 210034, China}

\author{M.~D. Ding}
\affiliation{School of Astronomy and Space Science, Nanjing University, Nanjing 210023, China}
\affiliation{Key Laboratory for Modern Astronomy and Astrophysics (Nanjing University), Ministry of Education, Nanjing 210023, China}

\author{Mats Carlsson}
\affiliation{Rosseland Centre for Solar Physics, University of Oslo, P.O. Box 1029 Blindern, NO-0315 Oslo, Norway}
\affiliation{Institute of Theoretical Astrophysics, University of Oslo, P.O. Box 1029 Blindern, NO-0315 Oslo, Norway}

\email{jiehong@nju.edu.cn,dmd@nju.edu.cn}

\begin{abstract}
The solar Ly$\alpha$ line is the strongest line in the ultraviolet waveband, and is greatly enhanced during solar flares. Here we present radiative hydrodynamic simulations of solar flares under different heating models, and calculate the response of this line taking into account non-equilibrium ionization of hydrogen and partial frequency redistribution. We find that in non-thermal heating models, the Ly$\alpha$ line can show a red or blue asymmetry corresponding to the chromospheric evaporation or condensation, respectively. The asymmetry may change from red to blue if the electron beam flux is large enough to produce a significant chromospheric condensation region. In the Ly$\alpha$ intensity lightcurve, there appears a dip when the change of asymmetry occurs. In thermal models, the Ly$\alpha$ line intensity peaks quickly and then falls, and the profile has an overall red asymmetry, which is similar to the profiles from heating by a soft electron beam. The Ly$\alpha$ profile shows a single red peak at the end of thermal heating, and the whole line is formed in a very small height range.
\end{abstract}


\section{Introduction}
The Ly$\alpha$ line is generated by transitions between the first two energy levels of the hydrogen atom. This line is the most intensive line in the solar ultraviolet waveband \citep{2001curdt}, due to the large abundance of hydrogen in the solar atmosphere. Since the first rocket flight observations of the Sun with the Ly$\alpha$ line \citep{1971gabriel,1975samain}, there has been a number of space instruments for spectroscopic and imaging observations of the Sun in recent years \citep{1995wilhelm,2012kano,2012kobayashi,2012woods,2016vourlidas}. 

Observations have shown that the Ly$\alpha$ line has responses to a great variety of solar activities from the chromosphere to the corona, including filaments/prominences \citep{2010curdt,2017chintzoglou,2018susino}, type-II spicules \citep{2018chintzoglou}, oscillations \citep{2011van,2017ishikawab,2017miligan} and flares \citep{2009rubio,2016miligan}. In particular, the Ly$\alpha$ line is greatly enhanced during flares, and the bright Ly$\alpha$ emission is co-spatial with the hard X-ray sources \citep{2009rubio,2016miligan}. In addition, the scattering polarization of this line, which is sensitive to the Hanle effect, has provided a new potential to explore the magnetism of the solar chromosphere \citep{2017ishikawaa,2018trujillo}.

There has also been theoretical studies of this line in order to help interpret observational results. Semi-empirical models have shown that the Ly$\alpha$ line is optically thick in the quiet Sun. The line center is formed around the transition region while the line wings are formed in the mid-chromosphere \citep{1981vernazza}. \cite{2017schmit} studied the correlation of the Ly$\alpha$ line with the \ion{Mg}{2} h line which is formed in the mid-upper chromosphere. 
They found that for both observations and simulations, there is a good correlation between the core intensities of the two lines. However, the correlation between the other diagnostics of the two lines in observations does not agree well with that in simulations.
 As for solar flares, there have been some modeling attempts that found an abrupt temperature rise and a quick brightening of the Ly$\alpha$ flux \citep{2005allred,2011rubio,2012rubio}. These studies made use of the radiative hydrodynamics code RADYN \citep{1992carlsson,1995carlsson,1997carlsson,2002carlsson} which applies a truncated profile to mimic the partial frequency redistribution (PRD) effect of the Ly$\alpha$ line. \cite{2018brown} made a progress in the calculation of the line by considering full PRD with the RH code \citep{2001uitenbroek,2015pereira}, while making an assumption of statistical equilibrium. For the first time, they showed the temporal evolution of the Ly$\alpha$ line profiles during a flare, and found that the profile has a centrally reversed core in general. They also found that this line can be blueshifted, while an overall red asymmetry is likely to be observed due to a low instrumental resolution. Recently, \cite{2019druett} made calculations of this line using the HYDRO2GEN code \citep{2018druett} with a truncated profile to mimic the PRD effect, and found that the wing emissions of the Lyman lines can reflect macro-motions in the chromosphere. As \cite{2018brown} has mentioned, non-equilibrium effects are important in determining the Ly$\alpha$ line, it would be highly desirable to include both effects in the calculation of this line.

In this paper, we investigate the response of the Ly$\alpha$ line during a flare including both the non-equilibrium effects and the PRD effects. We also employ different flare heating models. In addition,  we aim to focus on the impulsive phase of a flare, while the previous studies mainly focused on the relaxation phase. We briefly introduce our method in Section~2. In Section~3, we present our results of the Ly$\alpha$ line in response to different flare heating models. Then we discuss several diagnostics of this line in Section~4, followed by the conclusions in Section~5.

\section{Method}
\subsection{Flare modeling}
We employ the one-dimensional radiative hydrodynamics code RADYN in our modeling of solar flares. RADYN can calculate the atmospheric response to particle beam heating during a flare \citep{2015allred,2018prochazka} by solving the hydrodynamic and radiative transfer equations implicitly \citep{1992carlsson}. A more detailed description of the code can be found in \cite{2015allred}.

\begin{figure}
\plotone{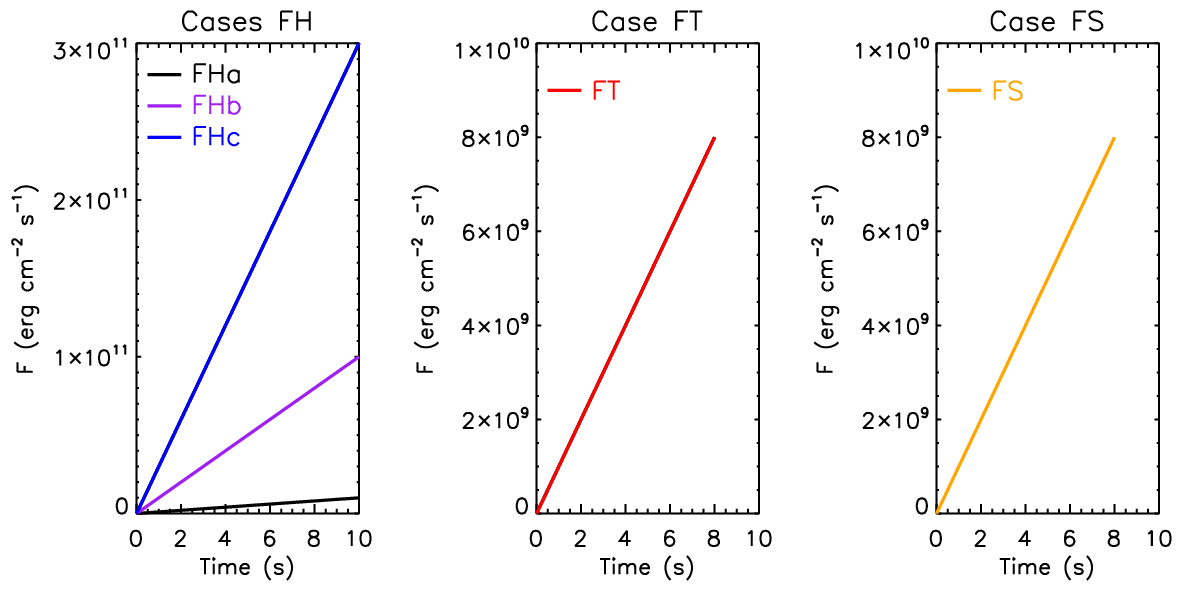}
\caption{Time evolution of the heating flux in the five simulation cases.}
\label{flux}
\end{figure}

The initial atmosphere has a quarter-circular loop structure with a 10 Mm length. We adopt the quiet-Sun model from our previous simulations \citep{2017hong,2018hong} which is based on the VAL3C atmosphere \citep{1981vernazza}. The initial temperature at the loop top is 10 MK. Here we consider two different heating mechanisms during a flare: non-thermal heating (through a beam of non-thermal electrons) and thermal heating (through direct plasma heating). For non-thermal flare heating, we assume a non-thermal electron beam with a spectral index of 3 and a low-energy cutoff of 25 keV. The energy flux of the beam rises linearly with time for a period of 10 s (Fig.~\ref{flux}). We consider three cases with different peak energy fluxes $F_{peak}$ (the value at the end of the simulation), labeled FHa, FHb and FHc respectively. Case FHa has a peak flux of $1\times 10^{10}$ erg cm$^{-2}$ s$^{-1}$. The peak flux of Case FHb is one order larger than that of Case FHa, and the peak flux of Case FHc is three times that of Case FHb. For thermal flare heating, we assume that the energy is deposited uniformly through the uppermost 8 Mm of the loop (Case FT). In Case FT, the total thermal heating rate at each time step is exactly the same as the total heating rate deposited by the electron beam in Case FHa (Fig.~\ref{flux}). 
However, due to a numerical instability, we only calculate the first 8 s in Case FT.
 In order to make a better comparison between thermal and non-thermal models, we also consider another case with a very soft non-thermal electron beam (Case FS). The electron beam in Case FS has a spectral index of 7 and a low-energy cutoff of 5 keV. Thus compared with the FH cases, Case FS has more energy deposited higher up in the atmosphere, which is similar to Case FT. The heating rate and the total duration of heating are exactly the same for Cases FS and FT (Fig.~\ref{flux}). All the parameters of the five models are summarized in Table~\ref{tab}. We run each of the five cases and save the simulation snapshots every 0.1 s.

\begin{table}[ht]
\centering
\caption{List of parameters of five flare models for simulation}
\begin{tabular}{ccccc}
\hline
Label & $F_{peak}$ (erg cm$^{-2}$ s$^{-1}$) & Total duration (s) & Spectral index & $E_{c}$ (keV) \\
\hline
FHa & $1\times 10^{10}$ & 10  & 3 & 25 \\
FHb & $1\times 10^{11}$ & 10  & 3 & 25 \\
FHc & $3\times 10^{11}$ & 10  & 3 & 25 \\
FT & $8\times 10^{9}$ & 8  & -- & -- \\
FS & $8\times 10^{9}$ & 8  & 7 & 5 \\
\hline
\end{tabular}
\label{tab}
\end{table}

\subsection{Calculation of the Ly$\alpha$ line}
In RADYN, all transitions are treated with complete frequency redistribution (CRD). For the Lyman lines where partial frequency redistribution (PRD) plays an important role, they are modeled with a Gaussian profile with Doppler broadening only \citep{2012leenaarts} to mimic the PRD effects.
 The RH code provides a good method to calculate spectral lines under PRD from a given atmosphere. However, different from RADYN, the original RH code assumes statistical equilibrium in the model atmosphere, which is not a good assumption in the impulsive phase of a flare, when the chromosphere is highly dynamic and the hydrogen ionization timescale is larger than the dynamic timescale of the atmosphere \citep{2002carlsson}.

We thus modify the RH code to take the hydrogen level populations and electron density directly from RADYN outputs and stop solving the statistical equilibrium equations. Doing so can ensure that, for each snapshot, the hydrogen populations reflect the instant conditions of non-equilibrium ionization. These populations are then taken into the PRD iterations to calculate the spectral line profiles. In this way, the non-equilibrium ionization of hydrogen and the PRD effects are both taken into account in the line profile calculations.

\begin{figure}
\epsscale{1.15}
\plotone{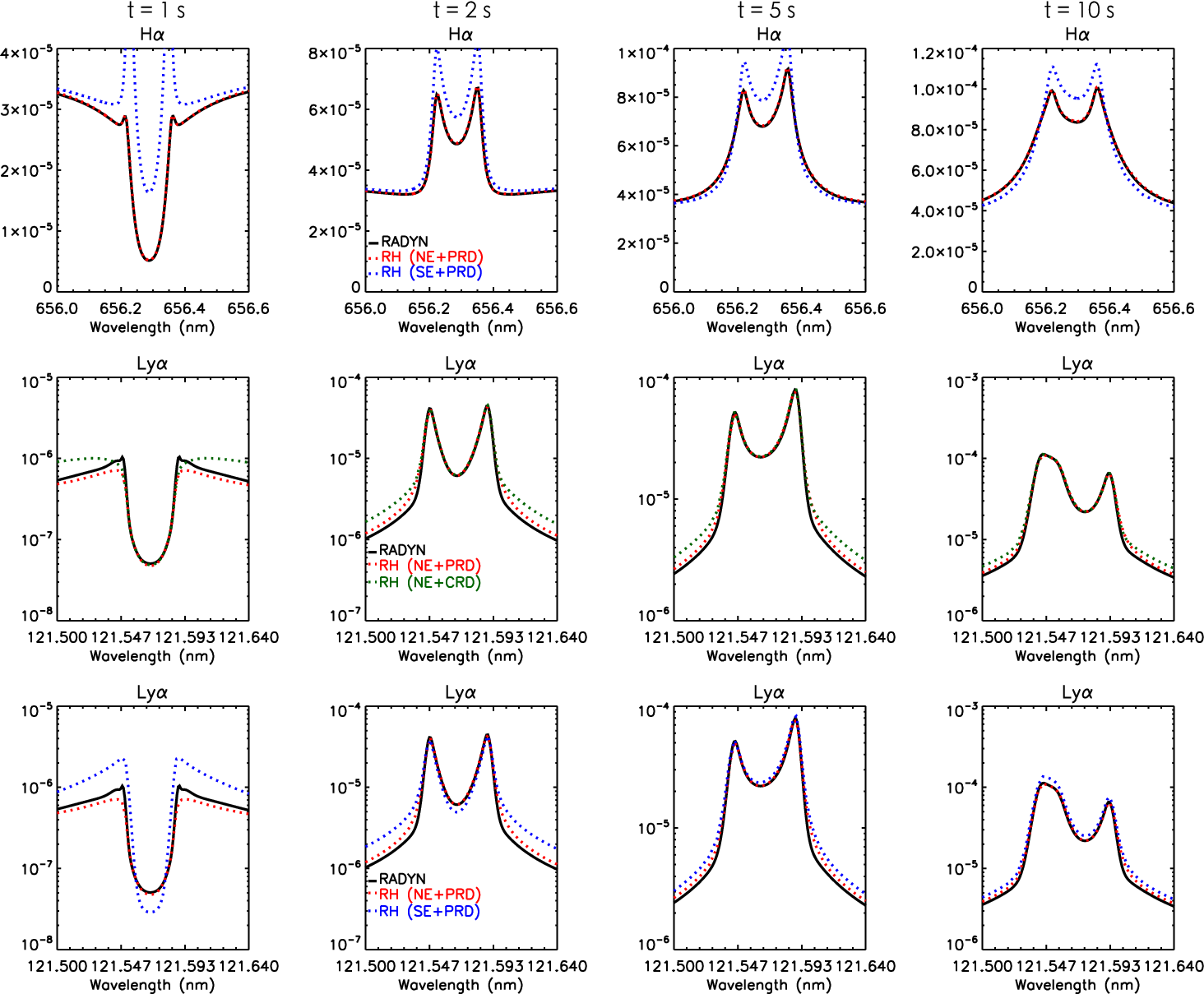}
\caption{The H$\alpha$ (top) and Ly$\alpha$ (middle and bottom) line profiles in Case FHb calculated under different assumptions. The black solid lines show results calculated directly from RADYN simulations, while the colored dotted lines show results calculated with the RH code. Red lines are under the assumption of non-equilibrium ionization (NE) and PRD; blue lines are under the assumption of statistical equilibrium (SE) and PRD; and green lines are under the assumption of NE and CRD. The Ly$\alpha$ profiles are in logarithmic scale.}
\label{ne}
\end{figure}

We compare the hydrogen line profiles in Case FHb calculated under different assumptions in Fig.~\ref{ne}. The top and the bottom rows show the non-equilibrium ionization (NE) effect on the H$\alpha$ and Ly$\alpha$ line profiles (see the difference between the red and blue lines), which is quite large during the impulsive phase of a flare. The NE effect has influences on both the line core and wings. The difference in intensity in the two assumptions of NE and SE can be more than twice in the line center of H$\alpha$ and the line wing of Ly$\alpha$. The middle row of Fig.~\ref{ne} shows the PRD effect on the Ly$\alpha$ line (see the difference between the red and green lines), which is significant only in the line wing. By contrast, for the H$\alpha$ line, the PRD effect does not influence the line profile obviously, so that the RH results converge to the RADYN results (see the red and black lines in the top row of Fig.~\ref{ne}). Note that for the Ly$\alpha$ line, the profile in RH still deviates from the profile in RADYN even under the same assumption like NE and CRD (see the difference between the green and black lines in the middle row of Fig.~\ref{ne}), since RADYN assumes a Doppler profile of this line while RH adopts a Voigt profile. The PRD effect on the Ly$\alpha$ line tends to be smaller when flare heating proceeds \citep{2018brown} but not vanishing, while the NE effect always works for the period of simulations. In the following, we calculate and present the Ly$\alpha$ line profiles under NE and PRD with the RH code.

\section{Results}
\subsection{Cases FHa, FHb and FHc}
\begin{figure}
\plotone{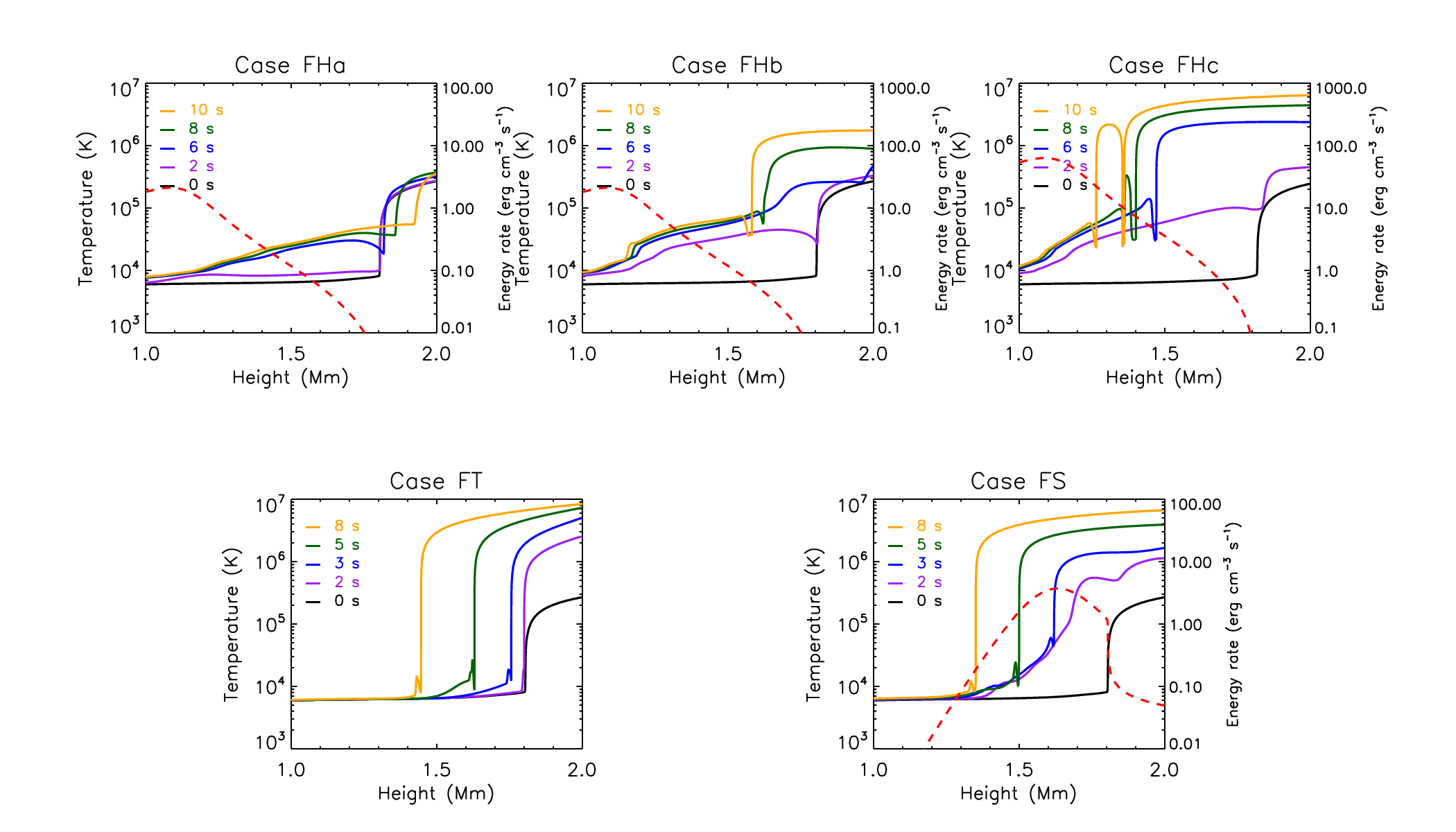}
\caption{Height distribution of temperature in the atmosphere, varying with the time of heating that is marked with different colors. The red dashed line denotes the beam heating rate in the atmosphere at 0.1 s.}
\label{temp}
\end{figure}

\begin{figure}
\plotone{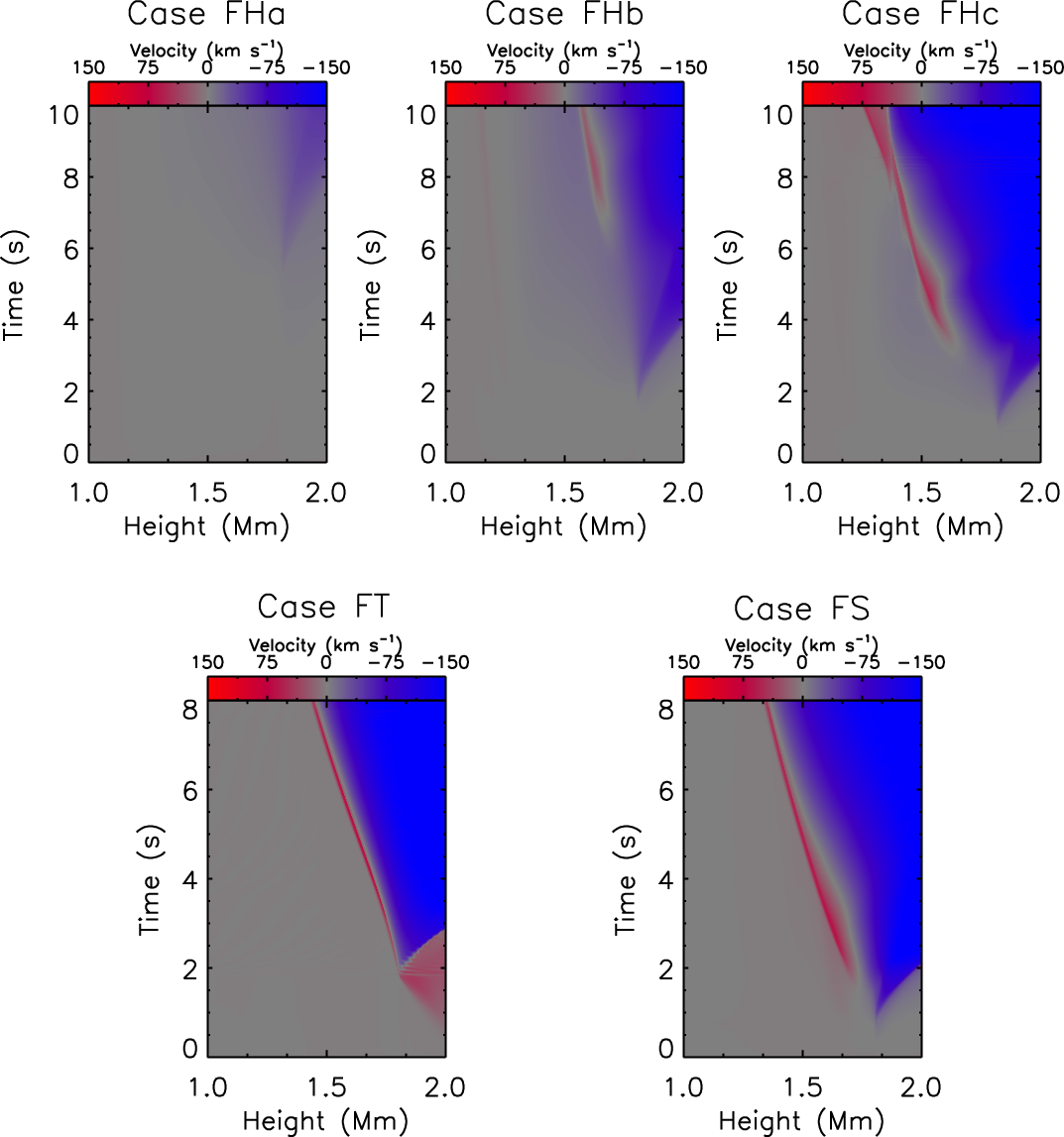}
\caption{Time evolution of the height distribution of vertical velocity where negative values are for upward motions (blueshifts) while positive values are for downward motions (redshifts).}
\label{vel}
\end{figure}

\begin{figure}
\plotone{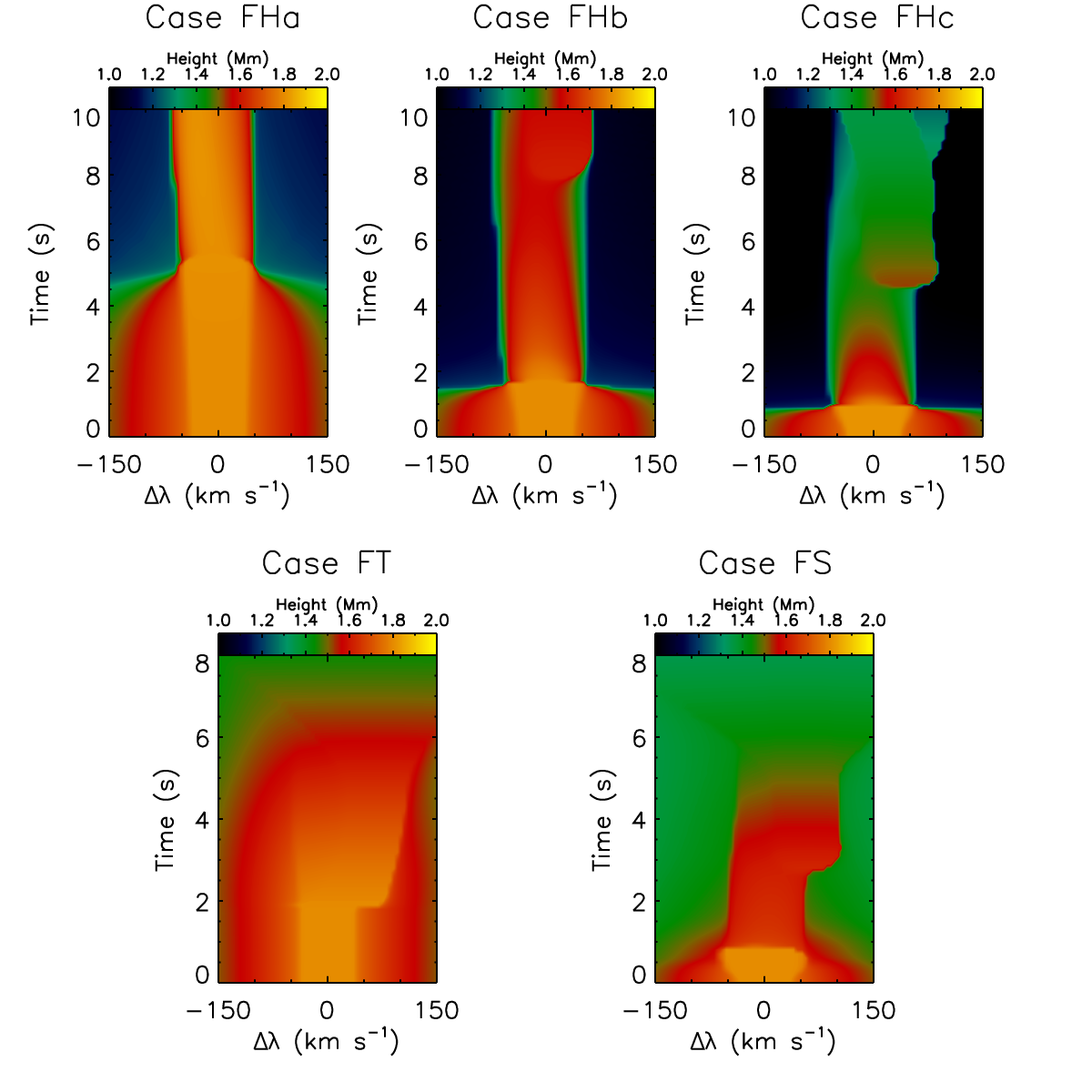}
\caption{Time evolution of the height where the optical depth reaches unity. The horizontal axis is in Doppler scale.}
\label{tau}
\end{figure}

\begin{figure}
\plotone{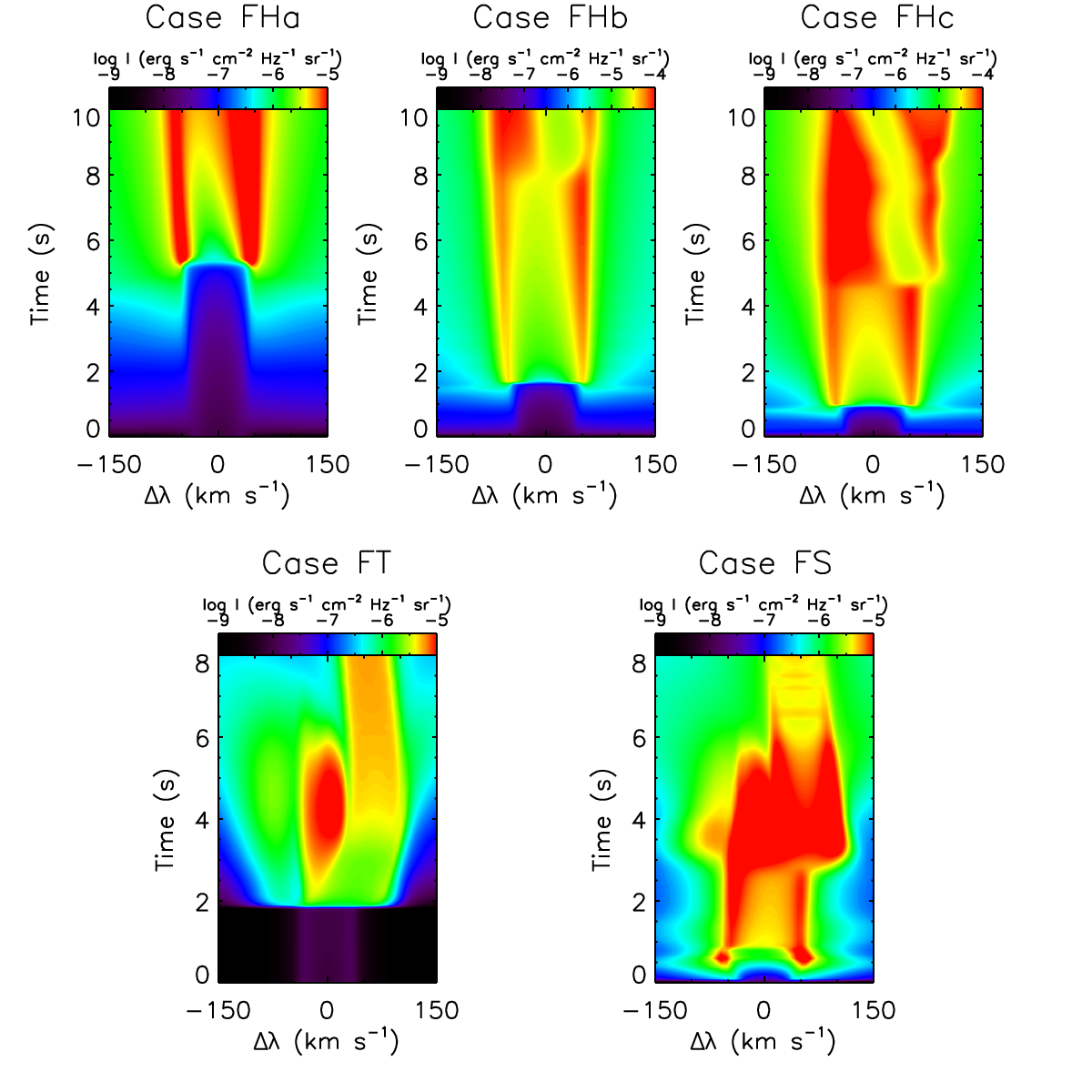}
\caption{Time evolution of the Ly$\alpha$ line profiles. Note that in each panel, the scale is different as denoted by each color bar. The horizontal axis is in Doppler scale. The abrupt asymmetry change is very clear in Cases FHb, FHc, and FS.}
\label{prof}
\end{figure}

Cases FHa, FHb and FHc represent a weak flare, an intermediate flare and a strong flare, respectively, heated by an electron beam with a hard spectrum. We show the time evolutions of the atmospheric parameters and the Ly$\alpha$ line profile in Fig.~\ref{temp}--\ref{prof}. In Case FHa, the beam heating rate has a peak at around 1.2 Mm, where most of the non-thermal electrons reach this height and deposit their energy locally (Fig.~\ref{temp}). As a result, the temperature in the chromosphere rises gradually and there forms a $1.7\times10^{4}$ K high-temperature plateau at 6 s. At this moment, the hydrogen atoms at ground level ($n=1$) start to get excited or ionized quickly. The high-temperature plateau then continues to rise, and the chromosphere expands and pushes the transition region upwards. Due to a relatively low heating rate, the chromosphere can not get heated to the coronal temperature, thus showing the typical features of gentle evaporation \citep{1985fisher}. The velocity of the evaporation can reach 40--50 km s$^{-1}$ at the end of the heating (Fig.~\ref{vel}).

\begin{figure}
\epsscale{1.25}
\plotone{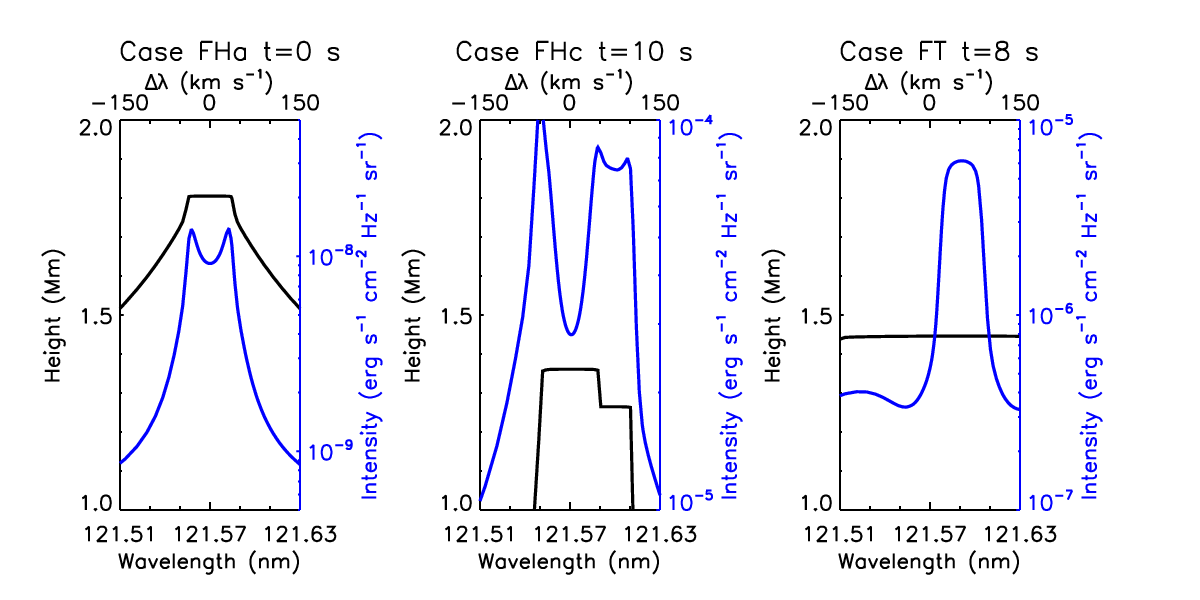}
\caption{The $\tau=1$ height (black) and the Ly$\alpha$ line profile (blue) in three cases at a certain time. Note that the horizontal axis is shown in both wavelength scale and Doppler scale with the exact same range.}
\label{evol2}
\end{figure}

The Ly$\alpha$ line is usually optically thick and there is a large span of the formation height from the line center to the line wings. In the initial atmosphere, the line center is formed near the transition region at around 1.8 Mm, while the line wings are formed in the mid-chromosphere (Fig.~\ref{evol2}). A central reversal is clearly seen in the Ly$\alpha$ line profile. When heating in the chromosphere proceeds and the hydrogen atoms get excited or ionized, the number density at the ground level is largely decreased, and the formation height of the line wings drops accordingly (Fig.~\ref{tau}). However, the formation height of the line center does not change too much since the opacity there is always very large, as reflected from the central reversal. The central reversal always exists no matter how the Ly$\alpha$ line intensity is enhanced by the flare heating (Fig.~\ref{prof}). During the chromospheric evaporation process, the line center is gradually blueshifted, corresponding to an upward velocity at the height where the line center is formed. The blueshift of the line center also causes a weak blue peak and a strong red peak, which we call a red asymmetry of the Ly$\alpha$ line profile.

In Case FHb, the electron beam flux and thus the heating rate are one order larger than that in Case FHa. 
The chromospheric temperature rises abruptly and a high-temperature plateau of more than $2\times 10^{4}$ K forms in the chromosphere (Fig.~\ref{temp}). After 8 s of heating, the atmosphere has reached the explosive phase when the non-thermal energy input can not be radiated away. The upper chromosphere is heated to a coronal temperature of nearly 1 MK, and shows an upward velocity of nearly 100 km s$^{-1}$ (Fig.~\ref{vel}), which are typical features of explosive chromospheric evaporation \citep{1985fisher,1999abbett}. At this time, there also appears to be a cool dense region with a downward-moving compression wave just below the newly-formed transition region at around 1.6 Mm (Fig.~\ref{temp}). This region is referred to as chromospheric condensation in momentum balance to the chromospheric evaporation \citep{1985fisherb,1999abbett,2018kowalski}. The condensation has a downward velocity of around 20 km s$^{-1}$, while the evaporation velocity has increased to 100 km s$^{-1}$ at 1.9 Mm.

When the chromosphere gets heated for about 2 s, a large proportion of the hydrogen atoms at the ground level is excited to higher levels or ionized, and the formation height of the Ly$\alpha$ line wings quickly drops (Fig.~\ref{tau}). During the first few seconds, the evaporation is manifested as a blueshift of the line center in the Ly$\alpha$ line profiles, resulting in a red asymmetry as in Case FHa (Fig.~\ref{prof}). When the upper chromosphere gets heated to the coronal temperature, the local number density of neutral hydrogen decreases quickly. As a result, the formation height, roughly corresponding to the height where $\tau=1$, of the Ly$\alpha$ line center decreases from the initial transition region at 1.8 Mm to the condensation region at around 1.6 Mm at the time of 8 s. The downward velocity of chromospheric condensation then induces a redshift of the Ly$\alpha$ line center, which leads to a stronger blue peak instead, called blue asymmetry here. The change from red asymmetry to blue asymmetry occurs just at the moment when the condensation region appears.

The behavior of the atmosphere and the Ly$\alpha$ line profiles in Case FHc are much more complicated than the previous two cases. The large electron beam flux can heat the chromosphere to more than $5\times10^{4}$ K after 2 s (Fig.~\ref{temp}), and the hydrogen atoms in the upper chromosphere are almost fully ionized. This case produces a strong explosive evaporation, and a significant condensation that initially forms at around 1.5 Mm. Similar to Case FHb, the Ly$\alpha$ line profile shows a change from red asymmetry to blue asymmetry during the flare evolution (Fig.~\ref{prof}). The condensation region gradually moves downwards as heating continues, reaching a velocity of around 40 km s$^{-1}$ at 6 s (Fig.~\ref{vel}). Interestingly, after 8 s of heating, there appears another cool dense region just below the former condensation region at 1.4 Mm. The plasma between these two cool regions is quickly heated to 1 MK, mainly due to an energy imbalance that the heating energy cannot be efficiently radiated away. The region of hot plasma in the mid-chromosphere tends to expand and partly reduces the downward velocity of the condensation region above, and the newly-formed condensation region below begins to move downwards. After 8 s, the $\tau=1$ height of the Ly$\alpha$ line center stays nearly in the upper condensation region, where the downward velocity gradually decreases with time (Fig.~\ref{tau}). However, the red wing of the line is formed just at the lower condensation region. Thus, the redshifted Ly$\alpha$ line center gradually moves back to the static position, and a second absorption feature (reversal) appears in the red wing (Fig.~\ref{evol2}). 

\subsection{Cases FT and FS}
Case FT is different from the previous cases in the heating model. In Cases FHa, FHb and FHc, energy is deposited in the chromosphere mainly through non-thermal electrons that propagate downwards along the flare loop from the loop top where magnetic reconnection is assumed to take place. In Case FT, however, the chromosphere gets heated through thermal conduction from an ad hoc heated layer that is the uppermost 8 Mm of the loop. The results show that, during the first 2 s, the lower corona is heated very quickly to a temperature of more than 10$^{6}$ K (Fig.~\ref{temp}). As heating continues, the upper chromosphere also gets heated and the transition region is gradually moving down. The heating rate by thermal conduction peaks around the transition region where the temperature gradient is very large. At 3 s, there appears to be a weak condensation just below the transition region, with a downward velocity of 50 km s$^{-1}$ (Fig.~\ref{vel}). Therefore, the originally symmetric profile of $\tau=1$ height is dragged to the red side (Fig.~\ref{tau}), and the line profile shows an obvious red asymmetry (Fig.~\ref{prof}). 
When the condensation region moves downward, it can shift the layer where $\tau=1$ to deeper layers at the Ly$\alpha$ line center, owing to a reduced opacity since the neutral hydrogen atoms above the condensation region are mostly ionized. The $\tau=1$ height of the line wing also moves down slightly since the change in opacity is relatively small compared with that in the line center. Eventually, at 8 s, the $\tau=1$ height is almost the same from the line center to the line wings. The Ly$\alpha$ line at this moment is thus formed in a very narrow height range, and the line profile is singly peaked with the peak being redshifted because of the condensation (Fig.~\ref{evol2}).

Case FS is similar to Cases FHa, FHb and FHc in the energy transport mechanism (non-thermal electron beam impact). However, a difference is that we adopt an electron beam with a softer energy spectrum in Case FS. Thus, at the beginning, most of the energy is deposited in the higher layers of the chromosphere around 1.6 Mm  (Fig.~\ref{temp}). As a consequence, at the first 2 s, the upper chromosphere is gradually heated to the coronal temperature. The electron density at this height increases by up to one order of magnitude due to an enhanced ionization of hydrogen. At 3 s, the model produces the chromospheric evaporation with a large upward velocity (nearly 100 km s$^{-1}$ at 1.85 Mm), accompanied by the condensation with a downward velocity (60 km s$^{-1}$) at around 1.6 Mm where the line center is formed (Fig.~\ref{vel}). As a result, the line center of Ly$\alpha$ is redshifted, leading to a stronger blue peak or a blue asymmetry. The temporal evolution of the atmosphere in Case FS after 3 s is very similar to that in Case FT. With the heating of the flare, the transition region is moving downwards and the $\tau=1$ height of the line center is also shifted to deeper layers (Fig.~\ref{tau}). The line profile is again complicated but finally evolves to a relatively simple one with a single, redshifted emission peak at 8 s (Fig.~\ref{prof}).

\section{Discussion}
\subsection{Contribution from non-thermal collisional rates}
\begin{figure}
\epsscale{1.15}
\plotone{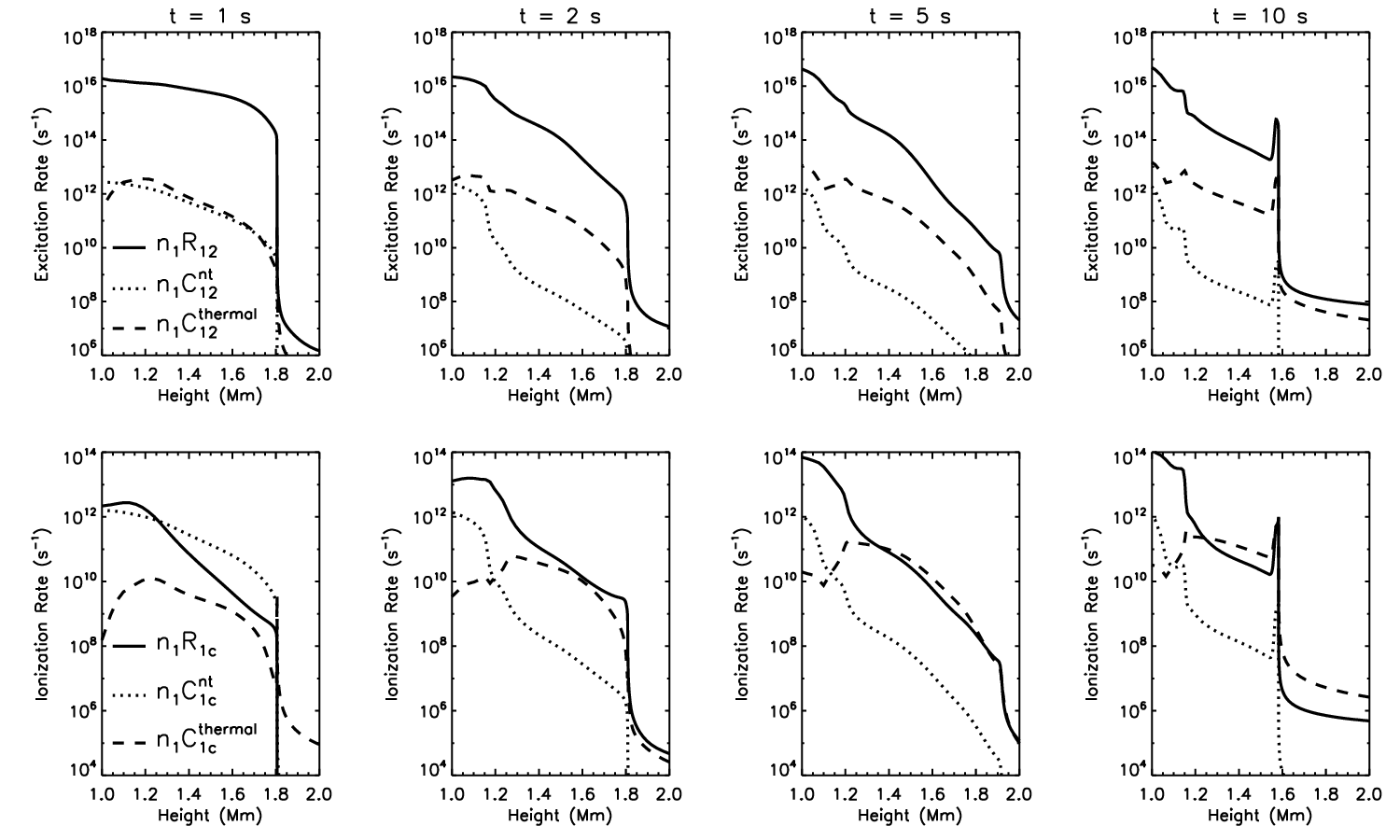}
\caption{Height distribution of the excitation rates from the hydrogen ground level to the second level, and ionization rates from the ground level. Radiative rates, non-thermal collisional rates and thermal collisional rates are shown as solid, dotted and dashed lines respectively.}
\label{rate}
\end{figure}

Besides the effect of non-equilibrium ionization of hydrogen, it is also interesting to evaluate the effects of non-thermal excitations and ionizations of hydrogen by the beam electrons. For a quantitative comparison, we plot the radiative and collisional rates from the hydrogen ground level to the second level (excitation) and that to the continuum level (ionization) as a function of height for Case FHb in Fig.~\ref{rate}. One can clearly see that for the $n=1\to 2$ excitation, the radiative rates always dominate the collisional rates by a factor of at least $10^{3}$. At the beginning of heating ($t=1$ s), the non-thermal collisional rate is about one order of magnitude larger than the thermal collisional rate at a height of 1.0 Mm. At chromospheric layers higher up, however, the non-thermal collisional rate is smaller than the thermal collisional rate. As the temperature of the chromosphere continues to rise, the thermal collisional rate begins to dominate the non-thermal collisional rate. For example, at $t=2$ s, the former is larger than the latter by a factor of around $10^{3}$ but is still much smaller than the radiative rate. For the ionization, the collisional rate can exceed the radiative rate in some part of the chromosphere at the beginning of heating ($t=1$ s). At this time, the non-thermal collision makes a large contribution to the ionization, which is larger than the thermal collision by a factor of at least $10^{2}$. The non-thermal collisional ionization keeps over the thermal one in the mid-chromosphere at the height of around 1.0 Mm during the whole simulation period. However, the thermal rate dominates the non-thermal one in the upper chromosphere after a short time of $t=2$ s. We conclude that for Case FHb, non-thermal excitations and ionizations from beam electrons do play a dominant role at the beginning of heating. When flare heating induces a drastic temperature increase in the chromosphere, the thermal collisional rates are greatly enhanced so as to exceed the non-thermal collisional rates. Therefore, for most of the simulation time in Case FHb (after $t=2$ s), the contribution to populations from non-thermal collisions are not important at the height above 1.2 Mm where the Ly$\alpha$ line center and peaks are formed, which agrees with the results presented in \cite{2019druett}. The non-thermal ionization rate in Case FHb is similar to that in \cite{2015allred}, while the thermal rate in Case FHb is much larger, possibly because \cite{2015allred} chose a cooler atmosphere as the initial atmosphere.

\subsection{Line asymmetries}
One of the good diagnostics of the Ly$\alpha$ line is the line asymmetry, which is subject to plasma motions in the atmosphere. In our simulations, this line shows different patterns of line asymmetries during the flare evolution (Fig.~\ref{prof}). In Cases FHa, FHb and FHc, the Ly$\alpha$ line shows a central reversal for most of the time. The influence of chromospheric evaporation and condensation on this line is to alter the position of the line center. As a result, the red asymmetry of the Ly$\alpha$ line (a stronger peak in the red wing) does not necessarily mean downward motions in the atmosphere, but rather upward motions at the formation height of the line center, and vice versa \citep{2018brown}. \cite{2018brown} showed that after 10 s of heating, the Ly$\alpha$ line profile shows a red asymmetry in Case F10D3 (similar to our Case FHa), while it shows a blue asymmetry in Case F11D3 (similar to our Case FHb). Our results are basically consistent with those of \cite{2018brown}. Thus, it seems that the effect of non-equilibrium ionization does not qualitatively influence the asymmetry of the Ly$\alpha$ line profile. In addition, \cite{2019druett} also reported enhanced red wing intensity of the Ly$\alpha$ line profiles at the beginning of the impulsive phase.

More interesting is that the line asymmetry can change with time as revealed in Cases FHb and FHc, where the beam heating rate is large enough to cause an explosive evaporation that is accompanied by a significant chromospheric condensation region below. The change of line asymmetry occurs when the downward motion influences the formation height of the Ly$\alpha$ line center. Note that in Case FHa, there is no apparent downward velocity, and thus the Ly$\alpha$ line shows a red asymmetry throughout the whole time period of simulation. Interestingly, Case FS also shows an asymmetry change in the first 2 s, when heating in the upper chromosphere has caused downward motions. 

It is also interesting to compare the Ly$\alpha$ and H$\alpha$ line asymmetries as shown in Fig.~\ref{ne}. \cite{2015kuridze} have simulated the H$\alpha$ line in a flaring atmosphere and pointed out that the asymmetry of the H$\alpha$ line is also interpreted as the movement of the line center due to evaporation or condensation flows, and that there is also a change of the line asymmetry from red to blue. We find that the asymmetry of the Ly$\alpha$ line shows a good agreement with that of the H$\alpha$ line, only that the blue asymmetry in the Ly$\alpha$ line is more obvious.

The evolution of the Ly$\alpha$ line profile in Case FT shows a very different pattern. The line profile is much more complicated with multiple peaks, but an overall red asymmetry is still visible. The red asymmetry later evolves to an extreme case where the whole line profile shows a redshifted single peak. After 3 s, Case FS shows a similar evolution pattern of the line profiles to that in Case FT. 

In all the cases that the chromosphere is heated mainly through non-thermal electrons, the Ly$\alpha$ line shows a red asymmetry with a central reversal at the beginning due to a blueshifted line center. The red asymmetry then changes to a blue asymmetry if there is a chromospheric condensation region. 
While in Case FT where there is firstly a bulk heating in the corona and the chromosphere is then heated through thermal conduction, the Ly$\alpha$ line only shows a red asymmetry without a central reversal at an earlier time, and a redshifted single peak at a later time. The redshifted peak, however, corresponds to the downward motion in the chromospheric condensation region where the whole line is formed at this time. The different mass motions, i.e., chromospheric evaporation and condensation, in flare models with different heating ways and different energy deposit rates, can also be diagnosed by the \ion{Si}{4} line that is formed in the transition region in a similar way. Previous studies have shown that the optically thin \ion{Si}{4} line is blueshifted in response to upward motions during electron beam heating and redshifted in response to downward motions during thermal conduction heating \citep{2015rubio,2018polito}. However, a blueshifted \ion{Si}{4} line might corresponds to a blueshifted Ly$\alpha$ line center, which in turn shows a red asymmetry with a central reversal due to the large optical depth.

\subsection{Integrated line intensity}
\begin{figure}
\plotone{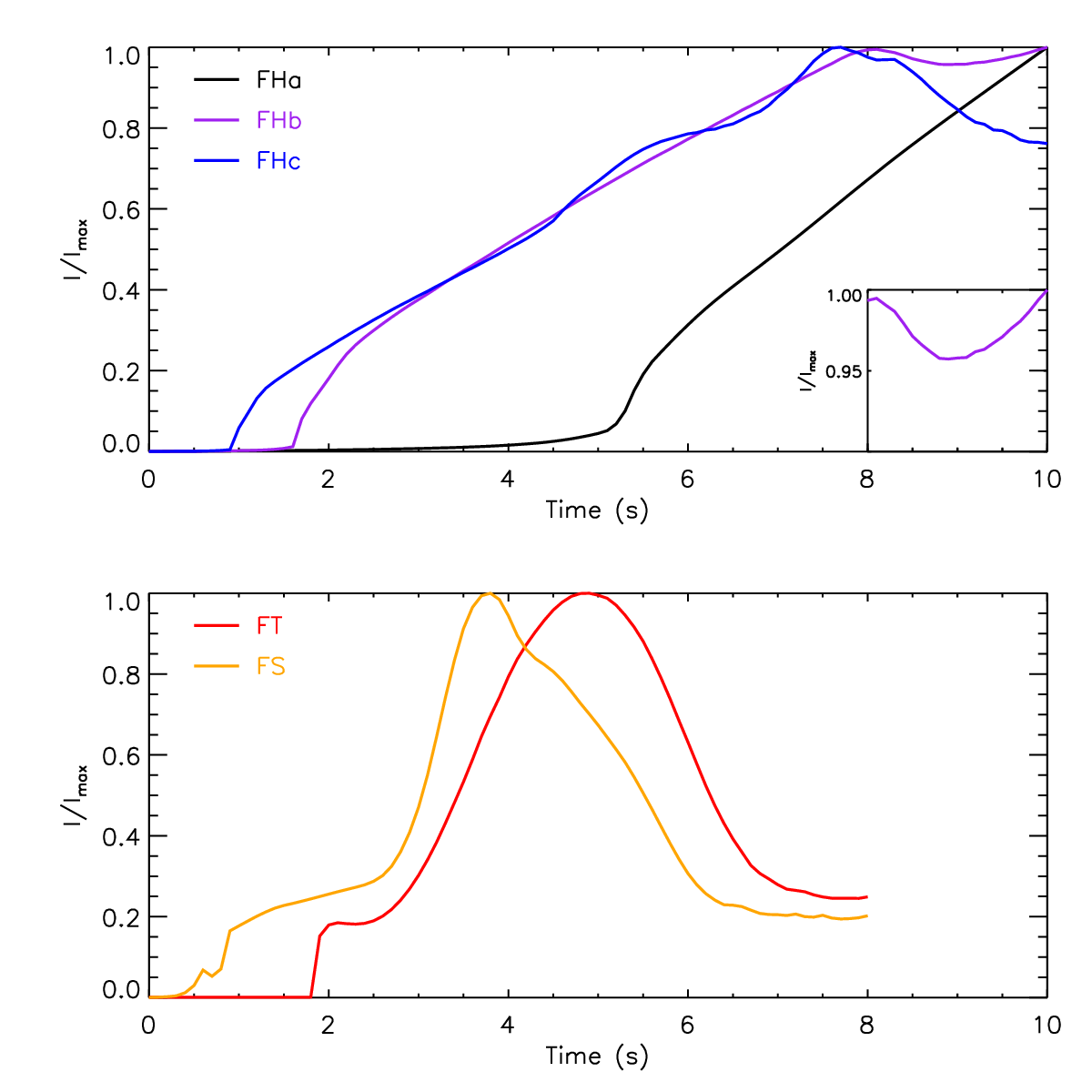}
\caption{Time evolution of the integrated intensity of the Ly$\alpha$ line. The intensity is normalized by the maximum intensity in each case. The integration range is from 121.50 nm to 121.65 nm. The lightcurve of Case FHb from 8 s to 10 s is also shown in the bottom right corner of the upper panel.}
\label{lc}
\end{figure}

We plot the time evolution of the integrated intensity of the Ly$\alpha$ line in Fig.~\ref{lc}. 
In Cases FHa, FHb and FHc, the intensity increases very gently at the beginning, and then grows more rapidly when a high-temperature plateau (around $1.8\times10^{4}$ K) forms in the chromosphere. The larger the beam flux is, the faster the chromosphere gets heated and the plateau forms, and the intensity rises rapidly in an earlier time. In Case FHc, the intensity begins to decrease after 8 s, when hydrogen atoms in the upper chromosphere are mostly excited or ionized. Although the intensity seems to continue increasing after 10 s in Cases FHa and FHb, it is expected to finally decrease at a later time (out of the time domain of our simulation).

The intensity evolution for Cases FT and FS is quite similar in spite of their different heating mechanisms. However, Case FT shows a longer latent time, during which the intensity has nearly no change, than Case FS. This is because the energy is deposited higher up in Case FT than in Case FS (see Fig.~\ref{temp}), and it needs some time for transporting the energy to the line formation region through thermal conduction. In both Cases FT and FS, the intensity reaches its maximum earlier than in Case FHa, simply because more energy is deposited in the line formation region in the former two cases. After the time of maximum intensity, the intensity in Cases FT and FS shows a clear and sharp decrease, even when the heating continues. 

In the cases showing a line asymmetry change, we also find an intensity dip in the Ly$\alpha$ lightcurve, coinciding with the asymmetry change. The intensity dip is quite obvious in Case FHb (at 9 s), while it is still visible in Case FS (at 1.2 s) but very difficult to recognize in Case FHc. The intensity dip for the Ly$\alpha$ line has also been shown in previous simulations \citep{2011rubio,2012rubio}.

\section{Conclusion}
In this paper, we explore the Ly$\alpha$ line profiles in response to different flare heating models. We solve the radiative hydrodynamics for both the heating models by a non-thermal electron beam and by thermal conduction using the RADYN code, and then calculate the Ly$\alpha$ line with the modified RH code including the NE effects. Our main results can be summarized as follows:

1. In the impulsive phase of a flare, the hydrogen ionization timescale is larger than the dynamic timescale of the atmosphere. Thus the NE effect has quite a large influence on the line profiles. For the Ly$\alpha$ line, the difference in the intensity at the line wings can be as large as a factor of two. The line center is also affected. However, it seems that the NE effect does not obviously affect the line asymmetry. The PRD effect also has a large influence on the Ly$\alpha$ line wings when compared with the CRD case, especially in an earlier time, while such an influence tends to be smaller in a later time of flare heating.

2. The Ly$\alpha$ line is usually optically thick with a central reversal at the beginning. In the beam heating model, non-thermal electrons can heat the chromosphere quickly, and the evaporation upflow is manifested as a blueshifted Ly$\alpha$ line center, producing a strong red peak, or a red asymmetry. Similarly, the condensation downflow is shown as a redshifted line center, producing a strong blue peak, or a blue asymmetry. Therefore, one should be cautious when relating the line asymmetries of the Ly$\alpha$ line to actual mass motions in the atmosphere \citep{2018brown}.

3. In the cases where the chromosphere is mainly subject to non-thermal electron beam heating (Cases FHa, FHb and FHc), the initial red asymmetry of the line could change to a blue asymmetry if the heating is strong enough to produce a significant chromospheric condensation region in the atmosphere. This corresponds to a dip in the time profile of the integrated line intensity of Ly$\alpha$.

4. In the case of thermal conduction heating (Case FT), the Ly$\alpha$ line shows an overall red asymmetry, and eventually a singly-peaked redshifted profile that is formed in a very narrow height region. If the electron beam is very soft (Case FS), the heating effect and the line profile is more similar to that in Case FT, in particular in a later time; however, a change of line asymmetry also appears in Case FS in the first 2 s, like in the Case FHb.

The results above suggest that the response of the Ly$\alpha$ line to different flare heating models is quite different. The change in line asymmetry and the dip in the lightcurve of the line might be specific features denoting heating through non-thermal electron beams. 
For heating through thermal conduction, the line profile shows an overall red asymmetry and it later evolves to a singly-peaked redshifted profile. However, in the case of heating by an electron beam with a very soft energy spectrum, the atmospheric response can present hybrid features, with the beam heating features at the beginning and the thermal conduction heating features at a later time.
Our results can provide diagnostics for future spectroscopic and/or imaging observations of this line with the new-generation high-resolution instruments like Solar Orbiter/EUI \citep{2014halain} and ASO-S/LST \citep{2016li}. 

\acknowledgments
We thank the referee for detailed and constructive suggestions that helped improve the paper. This work was supported by NSFC under grants 11733003, 11873095, 11533005 and 11961131002, and NKBRSF under grant 2014CB744203, and by the Research Council of Norway through its Centres of Excellence scheme, project number 262622. The authors thank ISSI and ISSI-BJ for the support to the team "Diagnosing heating mechanisms in solar flares through spectroscopic observations of flare ribbons". Y.L. is supported by CAS Pioneer Hundred Talents Program and XDA15052200, XDA15320301 and XDA15320103-03.

\clearpage

\end{document}